\documentclass[journal,onecolumn,draftcls,12pt]{IEEEtran}

\usepackage{times}
\usepackage{amsmath,dsfont, bm}
\usepackage{amssymb,amsthm}
\usepackage{epsfig,verbatim}
\usepackage{subfigure}
\usepackage{setspace}
\usepackage{color}
\usepackage{cite}
\usepackage{epstopdf}
\usepackage{graphics}
\usepackage{accents}
\usepackage{acronym}
\usepackage[bookmarks,breaklinks, colorlinks,pdfencoding=auto, psdextra]{hyperref}
\usepackage{soul}
\usepackage{enumitem}
\usepackage{mathrsfs}
\usepackage{bookmark}
\usepackage{algpseudocode}
\usepackage{algorithm}
\usepackage{float}
\usepackage{stackengine}
\def\delequal{\mathrel{\ensurestackMath{\stackon[1pt]{=}{\scriptstyle\Delta}}}}

\newtheorem*{theorem*}{Theorem}

\acrodef{dsa}[DSA]{dynamic spectrum access} 
\acrodef{tdma}[TDMA]{time-division multiple access}

\newcommand{\dsE}{\mathds{E}}
\newcommand{\Svec}{\mathbf{S}}

\newcommand{\mS}{\mathcal{S}}
\newcommand{\mX}{\mathcal{X}}
\newcommand{\mAac}{\mathcal{A}_{ac}}
\newcommand{\mAs}{\mathcal{A}_{s}}
\newcommand{\pac}{\pi_{ac}}
\newcommand{\ps}{\pi_{s}}
\newcommand{\RNum}[1]{\textup{\uppercase\expandafter{\romannumeral#1}}}

\begin{document}

\title{Deep Reinforcement Learning for Simultaneous Sensing and Channel Access in Cognitive Networks}

\vspace{-0.5cm}

\author{\IEEEauthorblockN{\vspace{-0.2cm} Yoel Bokobza, Ron Dabora and Kobi Cohen\\}\thanks{The authors are with the School of Electrical and Computer Engineering, Ben-Gurion University of the Negev, Be'er-Sheva, Israel (e-mail: yoelb@post.bgu.ac.il; ron@ee.bgu.ac.il; kobi.cohen10@gmail.com).\par
This work has been submitted to the IEEE for possible publication. Copyright may be transferred without notice, after which this version may no longer be accessible.\par
This work was supported by the Israel Science Foundation under Grant 584/20, and 5G WIN consortium.}
}

\markboth{October~2021}
{Shell \MakeLowercase{\textit{et al.}}: Bare Demo of IEEEtran.cls for IEEE Journals}

\maketitle

\begin{abstract}
We consider the problem of dynamic spectrum access (DSA) in cognitive wireless networks, where only partial observations are available to the users due to narrowband sensing and transmissions. The cognitive network consists of primary users (PUs) and a secondary user (SU), which operate in a time duplexing regime.
The traffic pattern for each PU is assumed to be unknown to the SU and is modeled as a finite-memory Markov chain. Since observations are partial, then both channel sensing and access actions affect the throughput. The objective is to maximize the SU's long-term throughput. To achieve this goal, we develop a novel algorithm that learns both access and sensing policies via deep Q-learning, dubbed Double Deep Q-network for Sensing and Access (DDQSA).
To the best of our knowledge, this is the first paper that solves both sensing and access policies for DSA via deep Q-learning.
Second, we analyze the optimal policy theoretically to validate the performance of DDQSA. Although the general DSA problem is P-SPACE hard, we derive the optimal policy explicitly for a common model of a cyclic user dynamics. Our results show that DDQSA learns a policy that implements both sensing and channel access, and significantly outperforms existing approaches.
\end{abstract}


\IEEEpeerreviewmaketitle

\section{Introduction}

The increasing demand for wireless communications and the limited availability of the electromagnetic spectrum have triggered the development
of efficient methods to increase the spectrum utilization in recent years. A main paradigm in this context is dynamic spectrum access (DSA), in which users monitor the spectrum to detect and access free channels for communications \cite{zhao2007survey}. There are two main approaches for implementing DSA in wireless networks: A centralized approach and a distributed approach. In centralized access management, there is a central network processor, which is a single point of contact for information sharing, whereas in the distributed management every node makes access decisions based only on its own observations without sharing information with other nodes.

In this paper, we focus on the design of distributed DSA for cognitive communication networks. In such networks, every user is designated as either a primary user (PU) or a secondary user (SU). When a PU requires a radio resource, a channel is allocated according to a predetermined resource allocation scheme which guarantees channel allocation to the PUs, while the SUs access the channel opportunistically and independently. To that aim, each SU independently monitors the wireless spectrum to identify free channels which are not being used by the PUs for communication. When properly designed, the incorporation of opportunistic SUs can achieve the desired overall increase in spectrum utilization \cite{akyildiz2006next}.
In practical implementations, due to bandwidth limitations in the sensing operation, an SU can sense only a part of the available spectrum (i.e., narrowband sensing), which implies that when operating in a distributed manner, access decisions are based on partial observations. For the purpose of this study we focus on a network that consists of a single SU which shares the spectrum resources with several PUs, where channel access is implemented in a \ac{tdma} manner, with fixed-length time slots. The transmissions of the PUs take place in frames whose length is a random variable (as it depends on the PU's incoming traffic, which is random) and may span several \ac{tdma} time slots.
For each PU, the random length of the transmitted frame is modeled as a finite-memory Markov chain, where different PUs may have different state transition probabilities for their corresponding chains. The SU does not have knowledge of the Markov chains of the PUs. As a result, at each time slot, based on its previous observations, the SU selects which channels to sense, and whenever it needs to transmit, it is allowed to select a single channel for transmission at the next time slot. Whenever the SU transmits on a channel that is not occupied by a PU, it receives an acknowledgment (ACK) signal indicating a successful transmission. Otherwise, a not-acknowledgment (NACK) signal is received denoting an unsuccessful transmission. The objective of the DSA algorithm in such a setup is to maximize the long-term rate of successful transmissions. 

\subsection{Related Work}
DSA has attracted a growing attention in past and more recent years, see e.g., \cite{zhao2007survey}, \cite{luong2019applications}. Related studies of DSA based on multi-armed bandit (MAB) formulations can be found in \cite{ahmad2009optimality, liu2010indexability, 6200864, 6362216, tekin2012approximately, 7006703, cohen2014restless, bagheri2015restless, gafni2020learning, gafni2021distributed}.
In the case of i.i.d channels, such that each channel is modeled as a 2-state Markov chain, representing the channel status as "busy" or "free", where in addition, the state transition probabilities are known a-priori at the SU, and under the assumption that when the channel is in a busy state it has a larger probability to remain in the busy state than to switch to the free state, the myopic policy has been proven to be optimal \cite{ahmad2009optimality}. 
In this strategy, the SU accesses the channel that will maximize the expected immediate reward without considering the effect of this action on future rewards.
While the myopic policy is easy to understand and simple to implement, it generally does not achieve optimal performance if one of the aforementioned assumptions is violated \cite{ahmad2009optimality}. Another algorithm that achieves optimal policy under the same optimality conditions of the myopic policy is the Whittle index algorithm \cite{liu2010indexability}. This algorithm has the advantage over the myopic algorithm in that it can lead to the derivation of good access policies even if the channels are not identically distributed. A major weakness of both the myopic algorithm and the Whittle index algorithm is that they are not applicable in scenarios in which the channels are correlated. Another major concern is that both the myopic policy and the Whittle index policy require full knowledge of the state transition probabilities, which is often unavailable in practical scenarios. 
This requirement has motivated the introduction of methods which can acquire an optimal policy approximately without requiring such a-priori knowledge. A major approach which is capable of achieving this goal is the reinforcement learning (RL) algorithm. RL is a class of machine learning algorithms, which can learn an optimal policy via interaction with the environment without knowledge of the system dynamics (such algorithms are also known as model-free algorithms) \cite[Ch.~1]{sutton2018reinforcement}.
Q-learning\cite{watkins1992q} is one of the most popular RL techniques which can directly learn the optimal policy online by estimating the optimal action-value function.
Early works which applied Q-learning to DSA used the classical tabular Q-learning method \cite{li2012dynamic}, \cite{venkatraman2010opportunistic}. However, it may be difficult to apply this method when the state space becomes large. 
This issue has motivated the combination of deep learning with RL, giving rise to the deep reinforcement learning (DRL) class of algorithms. These algorithms have attracted much attention in recent years due to their ability to approximate the action-value function for large state and action spaces. Recently, the work \cite{mnih2015human} proposed a DRL-based algorithm called deep Q-network (DQN), which is a combination of deep neural networks and the Q-learning algorithm. Recent studies that developed DRL-based algorithms for DSA problems can be found in \cite{wang2017deep, wang2018deep, nguyen2018deep, zhong2018actor, xu2018dealing, naparstek2019deep, zhong2019deep, xu2020application, zhang2020power, tan2020deep, 8962235}. In \cite{wang2017deep}, \cite{wang2018deep}, the authors applied DQN to the DSA problem, where it was assumed that at each time step, the SU can choose one channel to access and it receives an ACK/NACK signal as feedback from the accessed channel, based-on which the reward is computed. In \cite{wang2017deep}, \cite{wang2018deep}, the observations consist of the indices of the past accessed channels and the corresponding rewards, which are used as the inputs to the DQN algorithm.
In \cite{nguyen2018deep}, the authors assumed multi-band spectrum sensing without power limitations, which corresponds to a fully-observed scenario, and trained a DQN to select which channel to access in the next time step based on the current state of the entire spectrum.
In \cite{xu2018dealing}, \cite{xu2020application}, the authors used a deep recurrent Q-Network (DRQN), which is a combination of a DQN and a long short-term memory (LSTM) to derive the optimal access policy when observations are obtained using a fixed sensing pattern, in which at each time step the SU senses half of the available channels, such that a different half is sensed at each subsequent time step.
The LSTM layer in the DRQN algorithm uses past observations for the prediction of the state, which in turn, allows the agent to select a channel for accessing at the next time step. In \cite{zhong2018actor},\cite{zhong2019deep}, the authors used another DRL algorithm called deep actor-critic algorithm, which is a policy-based RL algorithm combined with a deep neural network. They compared their results with that of the algorithm in \cite{wang2018deep} and showed that their proposed algorithm achieves better performance. Power control aspects have been addressed in \cite{zhang2020power, tan2020deep}. In our recent work, we developed algorithmic solutions to reduce the size of DRL models when deploying in cheap hardware devices for DSA \cite{8962235}.

\subsection{Main Contributions}

We consider the DSA problem in cognitive wireless networks, consisting of multiple PUs and a single SU, as in \cite{wang2017deep, wang2018deep, nguyen2018deep, zhong2018actor, xu2018dealing, zhong2019deep, xu2020application, zhang2020power}. The PUs access the channel according to a predetermined policy, which guarantees channel allocation to every PU while the SU accesses the channel opportunistically.
As an SU can sense only a subset of the network's bandwidth (referred to as partial observations), this problem can be formulated as a partially observable Markov decision process (POMDP) \cite{wang2018deep}. In a case where the transition probabilities are known to the SU, an exact solution to this problem is P-SPACE hard and has an exponential computational complexity \cite{papadimitriou1987complexity}. In real-world models, the DSA problem is even harder to solve, as considered here, since the SU does not know the state probabilities of the PUs, nor which PU uses which channel. As a result, the SU does not know the state transition probabilities of the wireless network. Based on its past observations, the SU should select the best channel to access in the subsequent time step, such that the long-term throughput is maximized. To facilitate model-free learning we develop DRL-based algorithm to the design of the optimal policy. The novel and unique aspect of our approach is that as successful access decisions heavily rely on sensing, it is advantageous to train the agent to select {\em both} the channel to be sensed, as well as the channel to be accessed in the next time step. This is in contrast to previous works which designed DRL agents to maximize the throughput using predetermined sensing policies, or designed only access policies without a separate selection of channel sensing. Thus, the performance of these algorithms degrades under PU access policies in general. In this work, we develop a novel algorithm for a single agent that learns both access and sensing policies via deep Q-learning, dubbed Double Deep Q-network for Sensing and Access (DDQSA). For efficient learning, we employ a double deep Q-network (DDQN), which is a combination of double Q-learning \cite{hasselt2010double} and deep neural network, which facilitates learning from experience in an unknown environment with a large state space via interactions with the environment \cite{van2016deep}. To the best of our knowledge, this is the first paper that solves both sensing and access policies for DSA via deep Q-learning. Second, we analyze the optimal policy theoretically to validate the performance of DDQSA. Although the general DSA problem is known to be P-SPACE hard \cite{papadimitriou1987complexity}, we derive the first analytic development of the optimal policy for a common model of a cyclic PU dynamics. We tackle this challenge by exploiting the structure of the cyclic user dynamics to derive the optimal sensing policy that transfers this problem from a POMDP to a Markov decision problem (MDP). Then, we derive the corresponding optimal access policy using the optimality equations explicitly. Our results show that DDQSA learns a policy that implements both sensing and channel access, and achieves near-optimal performance. We further evaluate the throughput achieved by DDQSA under several PU access strategies, and compare the throughput obtained by DDQSA with those obtained by a DRL which uses deterministic sensing (i.e., a DRL which makes only access decisions with a predetermined sensing policy), and with an algorithm that performs an access without any sensing. The numerical simulations clearly demonstrate much better performance of DDQSA over existing approaches.

\subsection{Organization and Notations}

The rest of this paper is organized as follows: Section \RNum{2} details the network setup and assumptions; Section \RNum{3}, motivates and discusses the rationale for the selected DRL approach and details the proposed DDQSA algorithm. In Section \RNum{4}, we develop the optimal sensing and access policies for a network with a cyclic PU dynamics. These optimal policies serve as a benchmark for testing our algorithm. Section \RNum{5} reports simulation results, including a comparison with approaches proposed in previous works and with the optimal scheme (when possible). These results clearly demonstrate the advantages of the proposed approach over other approaches. Lastly, Section $\RNum{6}$ concludes this work.\par

We use $\mathds{N}$ to denote natural numbers, bold letters, e.g., $\mathbf{S}$ to denote vectors, and $\mathbf{S_i}$ denotes the $i$'th element in the vector $\mathbf{S}$, $i\geq0$. Calligraphic letters used to denote sets, e.g., $\mathcal{S}$, and the cardinality of a set is denoted by $|\cdot|$, e.g., $|\mathcal{S}|$ is the cardinality of the set $\mathcal{S}$.

\section{Problem Formulation}
We consider a wireless network with $N\in\mathds{N}$ channels, $K_p\in\mathds{N}$ PUs, $K_p\leq N$, and a single SU. 
We denote the $i$'th PU by $\mbox{pu}_i$, for $i\in\{0,...,K_p-1\}$, and abbreviate the $n$'th channel as $\mbox{ch}_n$, $n\in\{0,...,N-1\}$.
At each time step, the SU chooses a set of $L\in\mathds{N}$ channels, $L \le N$, for sensing to aggregate observations and predict which channel is the best channel for transmission at the next time step, in the sense of maximizing the long-term throughput.
For simplicity, we further assume that $L$ is a divisor of $N$.
The sensing outcome can be either "free", when the channel is not being accessed by any PU, or "busy" when the channel is being accessed by a PU.
The SU can base its prediction of the best channel for transmission in the next time step on all its past observations. If the SU has data for transmission, it transmits in the next time-step on the selected channel and receives feedback: 
If channel selection is successful, the SU receives an ACK message from the destination. Otherwise, the SU obtains a NACK signal which indicates that transmission has failed. Note that in practice, NACK can be determined by a timer and is not necessarily a response received from the destination.
For the subsequent discussion, we will assume that the SU transmits at each time step, and then, in Section \RNum{5}-C, we will show that this assumption can be easily removed with almost no effect on the algorithm's performance.

Following the natural intuition, we begin by formulating the problem as a POMDP with two policies: One policy for sensing and one for channel access. This formulation is quite intuitive and simplifies the explanations in the following sections. In Section \RNum{4}, we will show that this problem can be formulated as a single policy problem, and thus can be solved with a single agent.
Let $\mS=\{-1,1\}^N$ be the channel state space, 
where $'1'$ denotes that the channel is currently being used for transmissions and $'-1'$ denotes that the channel is free.
Let $\mathbf{s}^{(i)}\in \mathcal{S}$ be the $N$-length polar form representation of the integer $i\in\{0,1,...,2^N-1\}$, where the $j$'th element in $\mathbf{s}^{(i)}$, $\mathbf{s}^{(i)}_j$, represents the state of the $j$'th channel ($'1'$ if busy, and $'-1'$ if free), and let $\Svec(t)\in\mathcal{S}$ denote the state at time step $t\in\mathds{N}$. We define two action sets: 
$\mAs \delequal\{0,1,...,\frac{N}{L}-1\}$ is the action set for sensing, and $\mAac \delequal\{0,1,...,N-1\}$ is the action set for access. Accordingly, the set of $N$ channels is partitioned into sensing subsets, each sensing subset consists of $L$ channels, and the action  $A_s(t) = i \in \mAs$ denotes that the SU is sensing the channels $\{\mbox{ch}_m\}_{m=i\cdot L}^{(i+1)\cdot L-1}$ at time step $t\in\mathds{N}$. The action $A_{ac}(t)=j\in\mAac$ denotes that the SU decides to access $\mbox{ch}_j$ for transmission at time step $t+1$.
As the SU senses only a subset of $L$ channels in the network, the observations space, denoted by $\mX$, $\mX=\{-1,1\}^L$, is the set of all possible observation outcomes in the $L$ channels belonging to the currently observed subset.
The observation outcome in time $t\in\mathds{N}$ is denoted by $\mathbf{X}(t)\delequal\big[l, \big(x_0(t),x_1(t),...,x_{L-1}(t)\big)\big]$, where the first element  $\mathbf{X}_0(t)=l\in\{0,1,...,\frac{N}{L}-1\}$ indicates that at time $t$, the 
$l$'th subset of channels was sensed, i.e., $A_s(t)=l$, and $\big(x_0(t),x_1(t),...,x_{L-1}(t)\big)\in\mathcal{X}$, denotes the sensing outcome at time step $t\in\mathds{N}$, where the sensing outcome $x_i(t)=1$ implies that at time step $t$ the $i$'th channel of the $l$'th sensing subset is busy, and $x_i(t)=-1$ implies that it is free.
We define two policies: The sensing policy, denoted by $\ps$, and the access policy, denoted by $\pac$. Let $R(t)$ denote the reward at time step $t$. When a transmission is successful, i.e., when at time step $t$ the algorithm correctly selects a free channel for transmission at time $t+1$ (receives an ACK signal at time $t+1$), we set the reward to $R(t+1)=1$, to encourage the agent to access free channels. When the selected channel at time $t$ is busy at time $t+1$ (i.e, SU receives NACK signal), then  $R(t+1)=-1$, to discourage the agent from selecting a busy channel for transmission.

Our objective is to derive an RL-based algorithm to identify the pair of policies  $\ps^*$,  and $\pac^*$ that maximize the expected accumulated discounted reward over 
an infinite time horizon, i.e. \[ 
(\ps^*, \pac^*)=\operatornamewithlimits{\rm argmax}\limits_{\ps,\pac}\Big\{ \dsE_{\ps,\pac}\big\{\sum_{t=1}^{\infty}\gamma^{t-1}R(t+1)|\mathbf{X}(1)\big\}\Big\}, \]
for a discount factor $\gamma\in[0,1)$.

\section{The Proposed DRL-Based Algorithm for Channel Sensing and Access}

In this section, we describe the proposed Double Deep Q-network for Sensing and Access (DDQSA) algorithm. Prior to detailing the algorithm, we briefly review Q-learning and DQN to motivate our selection of this algorithmic approach.
\subsection{Q-Learning}
Q-learning is a model-free RL algorithm. When applied to an MDP, and under certain assumptions, this algorithm obtains the optimal policy in the sense of maximizing the expected accumulated discounted reward for any given initial state \cite[Ch.~6]{sutton2018reinforcement}.
The Q-learning algorithm is a value-based RL algorithm, which means that it computes the optimal action-value function for finding the optimal policy.
Let $\mathcal{A}$ denote the set of actions, $\mathcal{S}$ denote the set of states, and let $q_{\pi}(s,a)$, $s\in\mathcal{S}$, $a\in\mathcal{A}$, denote the action-value function, which is the expected accumulated discounted reward starting from state $s$, picking action $a$, and following policy $\pi$ afterwards. The term $\gamma\in[0,1)$, denotes the discount factor.
Because we consider the case of infinite time-horizon problem, then \cite[Ch.~3]{sutton2018reinforcement}
\[q_{\pi}(s,a) \delequal \dsE_{\pi}\left\{\sum_{t=1}^{t=\infty}\gamma^{t-1}R(t+1)|\mathbf{S}(t)=s, A(t)=a\right\}.\] 
The optimal policy $\pi^*$, is a policy that satisfies $q_*(s,a)\delequal q_{\pi^*}(s,a) \geq q_{\pi}(s,a)$ for any policy $\pi$ and for every possible state-action pair, $(s,a)\in\mathcal{S}\times\mathcal{A}$.
The optimal policy can be obtained easily from the optimal action-value function, $q_*(s,a)$, as $\pi^*(s) = \operatornamewithlimits{argmax}\limits_{a\in\mathcal{A}}\{q_*(s,a)\}$. 
The Q-learning algorithm iteratively estimates the optimal action-value function for each valid state-action pair in an online manner as follow: At each time step $t\in\mathds{N}$, the agent observes a state $s\in\mathcal{S}$, selects an action $a\in\mathcal{A}$, receives a reward $r$ for executing the selected action $a\in\mathcal{A}$, and observes the next state $s'\in\mathcal{S}$. 
Then, the estimation of the corresponding $q_*(s,a)$, referred to as the Q-value and denoted as $Q(s,a)$, is updated according to the update rule:
\[Q(s,a) \leftarrow Q(s,a) + \alpha \cdot \big(r+\gamma\cdot\max_{a'\in\mathcal{A}}\{Q(s',a')\} - Q(s,a)\big),\] for some $\alpha\in(0,1)$ referred as the learning rate. To explore various state-action pairs, the action $a$ is selected according to an $\epsilon$-greedy policy, meaning that most of the time the selected action maximizes the estimated optimal action-value function, whereas in the rest of the time the action is selected randomly from the set of all valid actions. Mathematically, the agent at state $s\in\mathcal{S}$, selects an action $a=\operatornamewithlimits{argmax}\limits_{a'\in\mathcal{A}}\{Q(s,a')\}$ with probability $1-\epsilon$, and a uniformly random action from all possible actions in state $s$, with probability $\epsilon$. According to \cite[Ch.~6]{sutton2018reinforcement}, this algorithm is proven to converge to $q_*(s,a)$ with probability 1 if all of the state-action pairs are visited infinitely often, and a variant of the usual stochastic approximation conditions is satisfied. In a general DSA setting, as considered here, the transition probabilities are unknown and only partial observations are available. As a result, convergence is not guaranteed theoretically.

\subsection{Deep Q-Network}
While Q-learning performs well when the action and state spaces are small and provably converges for MDP formulation, it turns out that for large state and action spaces, this algorithm is impractical.
There are two main reasons for Q-learning impracticality for large state and action spaces: 
The first reason is that in the Q-learning algorithm, the agent has to visit multiple states and select different actions in each state to learn the optimal Q-value, which requires a very extensive exploration and may result in a long learning time.
The second reason is that in the Q-learning algorithm, the agent has to store the Q-value for every state-action pair, which results in large storage requirements for large action and state spaces.
Recently, a class of DRL algorithms that combines Q-learning and deep neural networks, referred to as DQN, has been proposed. The role of the deep neural network in the DQN is to map observations and actions into their Q-values, which eliminates the need to store them in a table, thereby significantly reducing the storage requirement for large action and state spaces. Furthermore, the deep neural network has the ability to extract features from previous observations in order to infer the Q-value of observations that have not yet been observed \cite{nielsen2015neural}. This capability does not exist when using tabular methods, as when using tables, each state and action pair has to be visited to estimate the corresponding Q-value.

It should be noted that DQN-based learning is not guaranteed to converge to the optimal solution theoretically, even for problems which can be formulated as an MDP. In practice, however, it achieves very good performance even in various POMDP models with infinitely large state space. For example, the work of \cite{mnih2015human}, developed a DQN algorithm for teaching an agent how to play Atari games directly from screen images, and achieved very good performance in various Atari games.
Nevertheless, in the POMDP framework of the DSA problem, DQN suffers from performance degradation due to very partial observations which do not provide sufficient information about the entire channel states.

To cope with this problem, our proposed approach is to select observations wisely in the algorithm design, such that considering the state as a combination of a sufficient number of past observations, the algorithm can better infer about the actual system state, and find a (near-)optimal access policy to maximize the throughput. The novelty of our approach is the {\em joint learning} of {\em efficient sensing and access policies} via online learning by implementing a modified version of DQN, known as DDQN.

\subsection{The Proposed DDQSA Algorithm}

We start by showing that the DSA problem defined in Section \RNum{2} can be formulated as a single agent problem with a single policy for both sensing and access. Then, we introduce the DDQSA algorithm. 

Due to the partial observations, we maintain a history vector consisting of $H\in\mathds{N}$ most recent past observations to facilitate extracting more information about the state of the channel. We define the history-observations space as $\mathcal{O}_H = \{-1,0,1\}^{N\cdot H}$, where $\mathbf{O^{H}}(t) = [\mathbf{O}(t-H+1), \mathbf{O}(t-H+2), ..., \mathbf{O}(t)]\in \mathcal{O}_H$.
$\mathbf{O}(t)$ is a length $N$ vector that represents the observation outcomes at time step $t\in\mathds{N}$, where $\mathbf{O}_i(t) = 1$ denotes that  $\mbox{ch}_i$ was sensed at time step $t$ and was found to be busy, $\mathbf{O}_i(t) = -1$ denotes that  $\mbox{ch}_i$ was sensed at time step $t$ and was found to be free, and $\mathbf{O}_i(t)=0$ denotes that  $\mbox{ch}_i$ was not sensed at time step $t$.
The extended action space which facilitates selection of both sensing and access actions is denoted by $\mathcal{A}_{ex} = \{0,1,...,\frac{N^2}{L} - 1\}$. At each time step, the agent picks an action $a(t)\in \mathcal{A}_{ex}$, where $a(t) = i$, means that at time step $t$, the agent senses the channels $\{\mbox{ch}_m\}_{m=\lfloor\frac{i}{N}\rfloor\cdot L}^{(\lfloor\frac{i}{N}\rfloor+1)\cdot L-1}$ and transmits on channel $\mbox{ch}_{i\pmod{N}}$ at the next time step.

DDQSA utilizes the DDQN architecture originally proposed in \cite{van2016deep}, which combines double Q-learning and deep neural networks. Note that in the standard DQN, at the update step, a maximization operation is used for both selecting an action that maximizes the estimated Q-value and at the same time evaluating this Q-value with the selected action. This usually results in an overestimation of the Q-values, which causes performance degradation \cite{hasselt2010double}. In \cite{van2016deep}, the authors proposed to use two neural networks, one for selecting an action and the other for estimating the Q-value associated with the selected action, which was shown to achieve better performance.
In DDQSA, we use the observations history $\mathbf{O}^H(t)$ as the state of the wireless network which is used directly as an input to the DDQN. The output layer of the network, consisting of $|\mathcal{A}_{ex}|$ neurons where the $i$'th neuron in the output layer, $i\in\mathcal{A}_{ex}$, represents an estimation of $q_*(\mathbf{O}^H(t),i)$.

Note that although the immediate reward $R(t+1)$ is the same for any action $a(t)$ with the same value of $a(t) \pmod{N}$, the next observation will be different, thus the target, $R(t+1)+\gamma \cdot \max\limits_{a'\in\mathcal{A}_{ex}}\big\{Q\big(\mathbf{O}^H(t+1),a'\big)\big\}$ will be different and this would result in the agent learning a better joint sensing and access policy over time, where the learned access policy matches the corresponding sensing policy.
Following this formulation, the objective can be stated as finding the optimal policy $\pi^*$ such that \[\pi^*=\operatornamewithlimits{argmax}\limits_{\pi}\Big\{ \dsE_{\pi}\big\{\sum_{t=1}^{t=\infty}\gamma^{t-1}R(t+1)|\mathbf{O}^H(1)\big\}\Big\}.\]

From this policy we can obtain both sensing and access policies by setting $\pi_s(\mathbf{O}^H) = \lfloor\frac{\pi(\mathbf{O}^H)}{N}\rfloor\in \mathcal{A}_s$, and $\pi_{ac}(\mathbf{O}^H) = \pi(\mathbf{O}^H)\pmod{N}\in \mathcal{A}_{ac}$ for any history observation outcome $\mathbf{O}^H\in\mathcal{O}_H$.
To balance between exploration and exploitation, we used the $\epsilon$-greedy policy: At each time step $t\in\mathds{N}$, the agent selects an action
$a(t) =\operatornamewithlimits{argmax}\limits_{a'\in\mathcal{A}_{ex}}\big\{Q(\mathbf{O}^H(t),a')\big\}$ with probability $1-\epsilon$, and selects a random action uniformly among all actions with probability $\epsilon$.

\subsection{Pseudocode of DDQSA}

Let $\mathcal{D}$ denotes the replay buffer \cite{mnih2015human} and $\pmb\theta$, $\pmb\theta^-$ denotes the policy network weights, and the target network weights, respectively. The steps of the proposed DDQSA algorithm are summarized in Algorithm 1 below:
\begin{algorithm}[H]
\caption{The DDQSA Algorithm for Optimizing Jointly Spectrum Sensing and Access}\label{alg:my_alg}
\begin{algorithmic}[1]
\State Initialize replay buffer $\mathcal{D}$ to capacity $C$, the mini-batch $\mathcal{M}_B$ with size $|\mathcal{M}_B|$, and the target network update rate $J$.
\State Initialize the policy network weights $\pmb{\theta}$ randomly.
\State Initialize the target network weights $\pmb{\theta^-}\leftarrow\pmb{\theta}$.
\State Observe $\mathbf{O}^H(1)$.
\For{time step $t = 1,2,...$}
    \State Set $\epsilon = \frac{1}{1+0.01\cdot t}$.
    \State $\epsilon$-greedy: $a(t) = \begin{cases}
                            \operatornamewithlimits{argmax}\limits_{a'\in\mathcal{A}_{ex}}\big\{Q\big(\mathbf{O}^H(t),a'\big)\big\} & \text{w.p. $1-\epsilon$}, \\
                            \text{random action, } a\in \mathcal{A}_{ex} & \text{w.p. $\epsilon$}.
\end{cases}$
\State Execute actions:\newline
        \hspace*{5em} $a_s(t)=\lfloor\frac{a(t)}{N}\rfloor$; 
        $a_{ac}(t)=a(t)\pmod{N}$.
\State Obtain the reward $R(t+1)$, and observe the next state $\mathbf{O}^H(t+1)$.
\State Store the tuple $\big(\mathbf{O}^H(t)$, $a(t)$, $R(t+1)$, $\mathbf{O}^H(t+1)\big)$ in $\mathcal{D}$.
\State Sample a mini-batch $\mathcal{M}_B\delequal\{(\mathbf{o}_i, a_i, r_i, \mathbf{o}'_i)\}|1\leq i \leq |\mathcal{M}_B|\}$ randomly from $\mathcal{D}$.
\State Set target $y_i = r_i + \gamma \cdot Q(\mathbf{o}'_i, \operatornamewithlimits{argmax}\limits_{a'\in\mathcal{A}_{ex}}\big\{Q(\mathbf{o}'_i,a' ;\pmb{\theta})\big\};\pmb{\theta^-})$ and perform batch training with \hspace*{1.1em} inputs $\mathbf{o}_i$, and outputs $y_i$, using all $(\mathbf{o}_i, a_i, r_i, \mathbf{o}'_i)\in\mathcal{M}_B$.
\State Every $J$ iterations set $\pmb\theta^- \leftarrow \pmb\theta$.
\EndFor
\end{algorithmic}
\end{algorithm}

\section{Developing the Optimal Sensing and Access Policies for a Network with a Cyclic PU Dynamics}

In this section, we analyze the optimal policy theoretically to validate the performance of DDQSA. Although the general DSA problem is known to be P-SPACE hard \cite{papadimitriou1987complexity}, we derive the first structured solution of the optimal sensing and access policies analytically for the common model of a cyclic PU dynamics. In the next section, we will demonstrate that DDQSA indeed achieves a throughput which is very close to the optimal throughput.

Consider a network consisting of $N\geq 2$ channels, where $N$ is assumed to be even (note that the case of $N=2$ is trivial since both channels are sensed at each given time), the size of the sensing subset is $L=2$, and the number of PUs is $K_p=N-1$. The PUs access the channel according to the following rule: At each time step, the PUs either transmit at the same channel as in the previous time step with probability $P_{stay}$, or jointly switch to the adjacent channel to the right in a cyclic manner with probability $P_{switch}$, or jointly switch two channels to the right cyclically with probability $P_{Dswitch} = 1 - P_{stay} - P_{switch}$.
It follows from the description above that the state space of this network consists of four states, each corresponding to one possible location of the single free channel that can change its position with probabilities $P_{stay}$, $P_{switch}$, and $P_{Dswitch}$ according to the rules of the network.
Let $\mathcal{U}=\{0,1,,,,,N-1\}$ denote the set of possible free channels and let $U(t)\in\mathcal{U}$, such that $U(t)=i$ indicates that the free channel at time step $t\in\mathds{N}$ is the $i$'th channel, $\mbox{ch}_i$.
With these definitions, for any $s,s'\in\mathcal{U}$, the state transition probability is given by:
\begin{equation}
\label{transprobs}
\Pr(s'|s) = 
\begin{cases}
    P_{stay} & \text{if }s'=s, \\
    P_{switch} & \text{if }s'=s+1 \pmod N,\\
    P_{Dswitch} & \text{if }s'=s+2 \pmod N,\\
    0 & \text{otherwise}.
\end{cases}
\end{equation}
Setting the history length to $H=2$, we first analyze the possible observation outcomes and show that the optimal sensing policy $\pi^*_s$ results in a one-to-one mapping between the observations in the last two time steps and the current channel state $U(t)$, thereby making the state space fully observable:

\begin{itemize}
    \item 
When $\mathbf{X}(t) = [l,(-1,1)]$, $\mathbf{X}(t) = [l,(1,-1)]$ then, irrespective of $\mathbf{X}(t-1)$, since there is only one free channel, then it is obtained that the current state is $U(t)=l\cdot L$, and $U(t)=l\cdot L + 1$, respectively. 

\item Consider $\mathbf{X}(t) = [l,(1,-1)]$: We  conclude that $U(t)=l\cdot L + 1$, thus, $U(t+1)$ can be either $l\cdot L + 1$, $l\cdot L + 2 \pmod N$, or $l\cdot L + 3 \pmod N$. Hence, in the next time step, we must sense the pair of channels of subset $l+1 \pmod {\frac{N}{L}}$: 
\begin{itemize}
    \item If $\mathbf{X}(t+1) = [l+1 \pmod {\frac{N}{L}},(1,1)]$ then, $U(t+1)=l\cdot L+1$.
    \item If $\mathbf{X}(t+1) = [l+1 \pmod {\frac{N}{L}},(-1,1)]$, then $U(t+1)=l\cdot L + 2 \pmod N$. 
    \item If $\mathbf{X}(t+1)=[l+1 \pmod {\frac{N}{L}},(1,-1)]$, then $U(t+1)=l\cdot L + 3 \pmod N$.   
\end{itemize}

\item Consider $\mathbf{X}(t) = [l,(-1,1)]$: We  conclude that $U(t)=l\cdot L$, thus, $U(t+1)$ can be either $l\cdot L$, $l\cdot L+1$, or $l\cdot L+2 \pmod N$. Hence, in the next time step, we must sense the same pair of channels of subset $l$: 
\begin{itemize}
    \item If $\mathbf{X}(t+1) = [l,(-1,1)]$ then, $U(t+1)=l\cdot L$, 
    \item If $\mathbf{X}(t+1) = [l,(1,-1)]$, then $U(t+1)=l\cdot L+1$, 
    \item If $\mathbf{X}(t+1)=[l,(1,1)]$, then $U(t+1)=l\cdot L+2 \pmod N$.   
\end{itemize}

\item Assume that $\mathbf{X}(t-1) = [l,(-1,1)]$, and $\mathbf{X}(t) = [l,(1,1)]$. Then it is guaranteed that the free channel is now $U(t)=l\cdot L+2 \pmod N$. This state is equivalent to the partial observation $\mathbf{X}(t) = [l+1 \pmod {\frac{N}{L}},(-1,1)]$. Then, $U(t+1)$ is either $l\cdot L+2 \pmod N$, $l\cdot L+3 \pmod N$, or $l\cdot L+4 \pmod N$.
In this case,
\begin{itemize}
    \item If the sensing action is $A_s(t) = l+1 \pmod {\frac{N}{L}}$, then $\mathbf{X}(t+1)$ will be either $[l+1 \pmod {\frac{N}{L}},(-1,1)], [l+1 \pmod {\frac{N}{L}},(1,-1)]$, or $[l+1 \pmod {\frac{N}{L}},(1,1)]$ which has a one-to-one correspondence with $U(t+1)$ being equal to $l\cdot L+2 \pmod N, l\cdot L+3 \pmod N$, or $l\cdot L+4 \pmod N$, respectively. 
    \item If $A_s(t) = l$, then $\mathbf{X}(t+1)$ will be $[l,(1,1)]$ (if $N=4$, it will be either $[l,(1,1)]$, or $[l,(-1,1)]$). Then, if $\mathbf{X}(t+1)=[l,(1,1)]$ it follows that $U(t+1)$ may be $l\cdot L+2 \pmod N$ or $l\cdot L+3 \pmod N$, and we cannot know for sure the channel state.

\item We conclude that if the last two observations are $\mathbf{X}(t-1) = [l,(-1,1)]$ and $\mathbf{X}(t) = [l,(1,1)]$, the optimal sensing policy is to choose the sensing action $A_s(t) = l+1 \pmod {\frac{N}{L}}$ because it facilitates deterministic knowledge of the channel state at the next time step, $U(t+1)$.
\end{itemize}
\end{itemize}

\par Let $\mathcal{X}_{init}\delequal\big\{[l,(x_0,x_1)]\big|x_0=-1, \text{ or } x_1=-1 \text{ and } l\in\{0,1,...,\frac{N}{L}-1\}\big\}$ denote the initial set. By following similar reasoning, it follows that under the assumption of $\mathbf{X}(1)\in\mathcal{X}_{init}$, any consecutive pair of observations, $\big(\mathbf{X}(t-1), \mathbf{X}(t)\big)$, contains sufficient information to fully determine the network state $U(t)$. For example, Table \RNum{1} summarizes all of the possible observations for $N=4$ channels, their corresponding full network state, and the sensing policy, where $\pi_s\big(\mathbf{X}(t-1), \mathbf{X}(t)\big)=0$ denote that given the two observations at time steps $t-1$ and $t$, then  at time $t+1$ the SU will sense subset $0$ of the network channels, consisting of $\mbox{ch}_0$, and $\mbox{ch}_1$ , whereas $\pi_s\big(\mathbf{X}(t-1), \mathbf{X}(t)\big)=1$ denotes that given $\big(\mathbf{X}(t-1), \mathbf{X}(t)\big)$, at time step $t+1$ the SU will sense subset $1$ of the network channels consisting of $\mbox{ch}_2$, and $\mbox{ch}_3$.
\begin{table}
  \caption{Optimal sensing policy for $N=4$}
\begin{displaymath}
\begin{array}{|c|c|c|c|}
\hline
\mathbf{X}(t-1)& \mathbf{X}(t)  & U(t) & \pi_s\big(\mathbf{X}(t-1), \mathbf{X}(t)\big)\\ \hline
\text{don't care} & [0,(-1,1)] & 0 & 0\\ \hline
\text{don't care} & [0,(1,-1)] & 1 & 1 \\ \hline
\text{don't care} & [1,(-1,1)] & 2 & 1\\ \hline
\text{don't care} & [1,(1,-1)] & 3 & 0 \\ \hline
[0,(-1,1)] & [0,(1,1)] & 2 & 1 \\ \hline
[1,(-1,1)] & [1,(1,1)] & 0 & 0 \\ \hline
[1,(1,-1)] & [0,(1,1)] & 3 & 0 \\ \hline
[0,(1,-1)] & [1,(1,1)] & 1 & 1 \\ \hline
[1,(1,1)] & [1,(1,1)] & 1 & 1 \\ \hline
[0,(1,1)] & [0,(1,1)] & 3 & 0 \\ \hline
[0,(1,1)] & [1,(1,1)] & 0 & 0 \\ \hline
[1,(1,1)] & [0,(1,1)] & 2 & 1 \\ \hline
\end{array}
\end{displaymath}
\end{table}
Following this sensing policy, and under the assumption of $\mathbf{X}(1)\in \mathcal{X}_{init}$, the spectrum is fully observable at each time step, and hence this sensing policy is necessarily the optimal sensing policy.
As the state is fully observable, finding the best access policy can be formulated as an MDP problem.
According to the Bellman optimally equations \cite[Ch.~3]{sutton2018reinforcement}, the optimal policy $\pi^*_{ac}$ must satisfy $\pi^*_{ac}(s)=\operatornamewithlimits{argmax}\limits_{a\in\mathcal{A}_{ac}} \{q^*(s,a)\}$, where $q^*(s,a)$ is the optimal action-value function for $s\in\mathcal{U},a\in\mathcal{A}_{ac}$. Next, we characterize the optimal access policy: Letting $s'$ denote the next state, $U(t+1)$ and $v^*(s)$ denote the optimal access value function $\big(v^*(s)\delequal\operatornamewithlimits{max}\limits_{a\in \mathcal{A}_{ac}} \big\{q^*(s,a)\big\}\big)$, then $q^*(s,a)$ can be computed as:
\begin{align}
q^*(s,a) = &\sum_{r\in \{-1,1\},s'\in{\mathcal{U}}}\Pr(r,s'|s,a)\cdot \big(r+\gamma \cdot v^*(s')\big)\notag\\
= &\sum_{s'\in{\mathcal{U}}}\Pr(s'|s,a)\Big(\sum_{{r\in \{-1,1\}}}\Pr(r|s',s,a)\cdot \big(r+\gamma \cdot v^*(s')\big)\Big)\notag\\
{\overset{(a)}{=}} &\gamma\cdot\sum_{s'\in{\mathcal{U}}}\Pr(s'|s)\cdot v^*(s') + \sum_{s'\in{\mathcal{U}}}\Pr(s'|s)\cdot\Big(\sum_{{r\in \{-1,1\}}}r\cdot \Pr(r|s',a)\Big),\label{qstar}
\end{align}
where in step $(a)$, we used the fact that given the next state and the current action, the reward is fully determined, i.e., $\Pr(r|s',s,a) = \Pr(r|s',a)$, and that in the fully-observed case,  the action does not affect the probability of the next state given the previous state, i.e., $\Pr(s'|s,a)=\Pr(s'|s)$. Note that in the considered setup $\Pr(r|s',a)$ is deterministic, i.e.,
\begin{equation}\label{indicator}
\Pr(r=1|s',a) = \mathds{1}(s'=a), \ \Pr(r=-1|s',a) = \mathds{1}(s'\neq a),
\end{equation}
where $\mathds{1}(\mathsf{A})$ stands for the indicator function of the event $\mathsf{A}$.
From (\ref{indicator}),
\begin{align}
\sum_{s'\in{\mathcal{U}}}\Pr(s'|s)\sum_{{r\in \{-1,1\}}}r\cdot \Pr(r|s',a) &= \Pr(s'=a|s) - \sum_{s'\neq a}\Pr(s'|s)\notag\\ &= \Pr(s'=a|s) - (1-\Pr(s'=a|s))\notag\\ &= 2\cdot \Pr(s'=a|s)-1\label{qstarCont}.
\end{align}
Plugging (\ref{qstarCont}) into (\ref{qstar}) we obtain the optimal access policy as:
\begin{align}
\pi^*_{ac}(s)&=\operatornamewithlimits{argmax}\limits_{a\in\mathcal{A}_{ac}} \big\{q^*(s,a)\big\}\notag \\ &= \operatornamewithlimits{argmax}\limits_{a\in\mathcal{A}_{ac}} \Big{\{}\gamma\cdot\sum_{s'\in{\mathcal{U}}}\Pr(s'|s)\cdot v^*(s')+\sum_{s'\in{\mathcal{U}}}\Pr(s'|s)\cdot\Big(\sum_{{r\in \{-1,1\}}}r\cdot \Pr(r|s',a)\Big)\Big{\}}\notag\\ &{\overset{(a)}{=}} \operatornamewithlimits{argmax}\limits_{a\in\mathcal{A}_{ac}} \{2\cdot \Pr(s'=a|s)-1\}\notag\\ & =\operatornamewithlimits{argmax}\limits_{a\in\mathcal{A}_{ac}}\{\Pr(s'=a|s)\}.\label{pistar}
\end{align}
where $(a)$ follows as the first summand is independent of $a\in\mathcal{A}_{ac}$. Then, letting $P_{max} =  \max\{P_{stay}, P_{switch}, P_{Dswitch}\}$, from (\ref{transprobs}) and (\ref{pistar}) we conclude that for $s\in\mathcal{U}$ we have:
\begin{equation} 
\pi^*_{ac}(s)= \operatornamewithlimits{argmax}\limits_{a\in\mathcal{A}_{ac}}\{\Pr(s'=a|s)\} = 
\begin{cases}
    s & \text{if }P_{stay} = P_{max} \\
    s+1 \pmod N & \text{if }P_{switch} = P_{max}\\
    s+2 \pmod N & \text{if }P_{Dswitch} = P_{max}\\
\end{cases}.
\end{equation}

The above analysis leads to two important insights: First, we conclude that if the observations can be selected such that the agent can infer the full network state (as in the case analyzed in this section), the discount factor $\gamma\in[0,1)$ can be set arbitrarily. Practically, the best option will be to set $\gamma=0$, since for example, in the Q-learning algorithm the update rule becomes simply $Q(s,a) \leftarrow (1-\alpha)\cdot Q(s,a) + \alpha \cdot r$. As a result, the algorithm converges faster due to the elimination of the unnecessary term $\alpha\cdot \max_{a'\in\mathcal{A}}\{Q(s',a')\}$ from the update rule. This follows as the term $\alpha\cdot \max_{a'\in\mathcal{A}}\{Q(s',a')\}$ includes an estimate of the function $q_*(s',.)$ which may be very different from its true value at the beginning of the learning process.
The second insight is that our algorithm requires $\gamma>0$ to converge to the maximal throughput, which makes the suggested problem formulation a non-degenerated RL problem in the sense that actions selected in order to maximize the accumulated future rewards and not only the immediate reward. The reason is that in general, the full network state cannot be determined from a finite number of past observations. Furthermore, even in that case, then at the beginning of the learning process, the sensing policy is arbitrary. Then, usually the agent cannot infer about the channel state at the beginning. This implies that the agent must consider the effect of selecting which channels to sense on future rewards thereby improving its sensing policy.

Let $suc_i$ denote the event of a successful transmission at time step $i\in\mathds{N}$ and define the throughput of the optimal policy for this scenario as $T\delequal \lim_{t\rightarrow \infty}\frac{\sum_{i=1}^t\mathds{1}(suc_i)}{t}$. Following the optimal sensing policy implies that the events $\{\mathds{1}(suc_i)\}_{i=1}^{\infty}$ are i.i.d random variables. Then, according to the weak law of large numbers \cite{tabak2011probability} we have:
\[T= \lim_{t\rightarrow \infty}\frac{\sum_{i=1}^t\mathds{1}(suc_i)}{t} = \mathds{E}[\mathds{1}(suc_i)] = \Pr(suc_i).\]
According to the optimal access policy, the SU accesses a free channel with probability $P_{max}$, which implies that $T=\Pr(suc_i)=P_{max}$. Note that even if $\mathbf{X}(1)\notin \mathcal{X}_{init}$, we may choose sensing and access actions randomly. In this case, the probability that $\mathbf{X}(t)\notin \mathcal{X}_{init}$ for any $t\in\mathds{N}$ is $0$. Therefore, the probability that there exists $t_0\in\mathds{N}$ such that $\mathbf{X}(t_0)\in \mathcal{X}_{init}$ is $1$. Once $\mathbf{X}(t_0)\in \mathcal{X}_{init}$ for some $t_0>0$, we can apply the optimal sensing and access policies $\forall t>t_0$ and asymptotically achieve the optimal throughput, because in this case,
\begin{align}
T &= \lim_{t\rightarrow \infty}\frac{\sum_{i=1}^t\mathds{1}(suc_i)}{t}\notag\\&= \lim_{t\rightarrow \infty}\frac{\sum_{i=1}^{t_0}\mathds{1}(suc_i)}{t} + \lim_{t\rightarrow \infty}\frac{\sum_{i=t_0+1}^t\mathds{1}(suc_i)}{t}\notag\\ 
&= \lim_{t\rightarrow \infty}\frac{\sum_{j=1}^{t-t_0}\mathds{1}(suc_{j+t_0})}{t} = P_{max}.\label{tau}
\end{align}
Finally, note that for $j\geq1$, $j+t_0\geq t_0+1$, i.e. the optimal sensing and access policies are followed. Therefore, we conclude that $\lim_{t\rightarrow \infty}\frac{\sum_{j=1}^{t-t_0}\mathds{1}(suc_{j+t_0})}{t}=P_{max}$, 
for any $\mathbf{X}(1)=\big[l,\big(x_0(1),x_1(1)\big)\big], l\in\{0,1,...,\frac{N}{L}-1\}, \big(x_0(1),x_1(1)\big)\in\mathcal{X}$.

\section{Experiments}
In this section, we report the outcomes of experiments carried out to test and evaluate the performance of the proposed DDQSA algorithm. DDQSA was implemented as described in Algorithm~\ref{alg:my_alg} in Section \RNum{4}, with two hidden layers of a fully connected deep neural network, where each layer consists of 128 neurons with the rectified linear unit (ReLU) activation function, $ReLU(x) = \max\{0,x\}$. The activation function for each neuron in the output layer is the linear activation function $f(x) = x$.
The $\epsilon$-greedy policy has been applied such that $\epsilon = \frac{1}{1+0.01\cdot t}$, i.e., $\epsilon$ decays over time.
At each time step, a mini-batch of 64 samples $\big(|\mathcal{M}_B|=64\big)$ from the replay buffer is uniformly sampled and used for training. 
The Adam algorithm \cite{kingma2014adam} is used as the optimizer with the mean-squared error (MSE) loss function.
We set the discount factor to $\gamma=0.8$, the learning rate is $\alpha=10^{-4}$, the replay buffer capacity is $C=30000$, and $J = 20$. We define the relative throughput $\rho(\tau) \delequal \frac{\eta(\tau)}{\eta_{bound}(\tau)}$, $\tau\in\mathds{N}$, where $\eta(\tau)$ is the number of successful transmissions in the range of time steps beginning from $(\tau-1)\cdot 100+1$ up to time step $\tau\cdot 100$ divided by 100, and $\eta_{bound}(\tau)$ is defined as the number of time steps in which at least one channel was free, in the range of time steps beginning from $(\tau-1)\cdot 100+1$ up to  $\tau\cdot 100$, divided by 100.
Thus, $\eta_{bound}(\tau)$ is an upper bound on the throughput of any DSA algorithm for the setup defined in Section \RNum{2}. In the following, $\rho(\tau)$ is used as the figure-of-merit for evaluating the performance of the different algorithms. 
\par
In the experiments, we compare the performance of DDQSA with that of three other algorithms with fixed sensing or access policies:
\begin{enumerate}
  \item Random Access: In this policy, the SU does not employ sensing, and at each time step selects randomly and uniformly a channel for accessing.
  \item Random Sensing: In this policy, the SU randomly selects a subset of channels to sense, and uses these observations to learn an access policy by employing a DDQN. 
  \item Alternating Sensing: In this policy, the SU senses each of the subsets of channels alternatingly (as in \cite{xu2020application}) and uses these observations to learn an access policy by applying a DDQN.
\end{enumerate}
The performance of the different algorithms was obtained by averaging the outcomes of 30 independent experiments for each algorithm at each scenario.
\subsection {Comparison with the Optimal Policy for the Cyclic PU Network}
\begin{figure}
  \centering
  \includegraphics[width=0.9\textwidth]{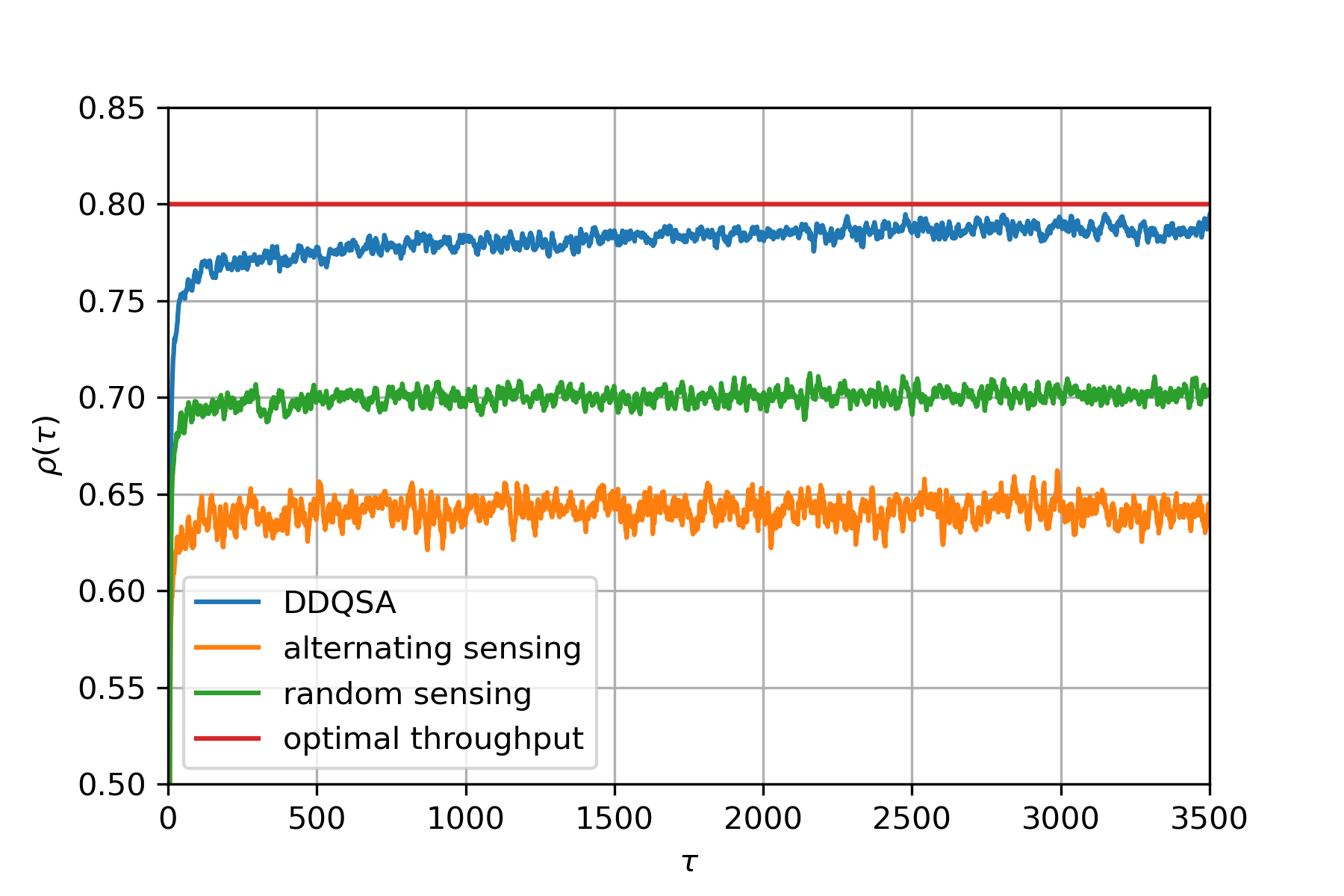}
  \caption{Relative throughputs for the cyclic PU network defined in Section \RNum{4}.}
  \label{fig:optimal}
\end{figure}

First, we consider the cyclic network defined in Section \RNum{4} with $N=4$ channels. In this network, $\eta_{bound}(\tau) = 1, \forall\tau\in\mathds{N}$, since there is always a single free channel at each time step, thus, $\rho(\tau) = \eta(\tau)$.
As obtained in Section \RNum{4}, the throughput of the optimal access and sensing policies for this network is $P_{max}$. 
In this experiment we set $P_{stay} = 0.1, P_{switch} = 0.1$, and $P_{Dswitch} = 0.8$, which implies that for $\tau\gg 1$, $\rho(\tau) = \eta(\tau) \approx P_{max} = P_{Dswitch} = 0.8$.
The throughput of the random access algorithm can be analytically obtained as $\rho_{RA}(\tau)=0.25$, which follows by noting that there are 4 channels, where at each time-step there is a single free channel.
Fig. \ref{fig:optimal} depicts the simulation results for this scenario. From Fig. \ref{fig:optimal}, we observe that the DDQSA algorithm performs well and asymptotically attains near-optimal performance (the throughput is numerically evaluated at approximately $0.79$), whereas under alternating sensing policy, the throughput is about $0.64$ and under the random sensing policy it is about $0.7$, both are highly sub-optimal.
We conclude that DDQSA is indeed capable of learning near-optimal joint sensing and access policy, thereby justifying  the rationale of our proposed approach.
\subsection {Experiment Results for Other General Scenarios}
We consider now a network consisting of $N=4$ channels, with $4$ PUs, $K_p=4$, observations subsets of size $L=2$, and a history length of $H=6$. The PU transmissions occur in frames which may span more than one time step according to the Markov model described in Section \RNum{2}:
For $\mbox{pu}_i$, $i\in{\{0,1,...,K_p-1\}}$, we set the maximal frame length to $l_i$, and let $\mathcal{L}_i = \{0,1,...,l_i\}$ denote the state space for $\mbox{pu}_i$. When $\mbox{pu}_i$ is at the $k$'th state of its frame we set its state to $m_i=k$. When $\mbox{pu}_i$ is not transmitting, referred to as idle, we set its state to $m_i=0$.

\begin{figure}[H]
  \centering
  \includegraphics[width=0.8\textwidth]{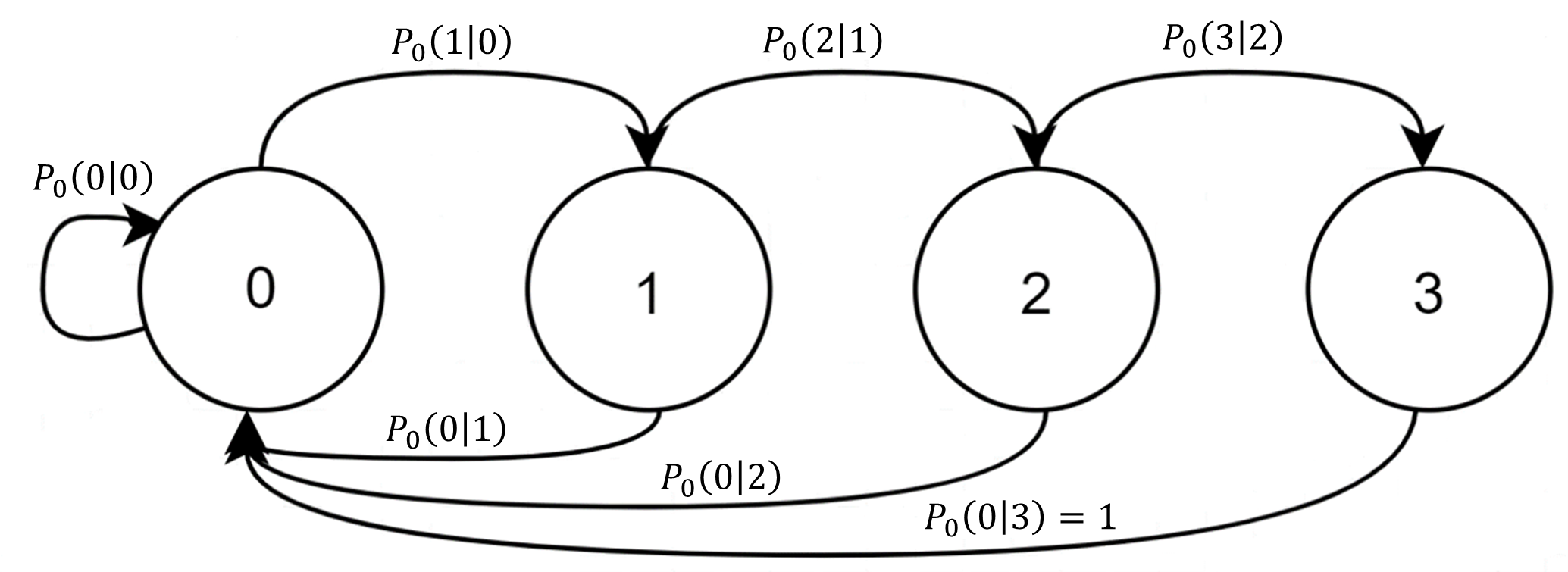}
  \caption{An illustration of the state transition diagram for $\mbox{pu}_0$.}
  \label{fig:markov_chain}
\end{figure}

Denote by $P_{i}(k|j)$, $j,k\in\mathcal{L}_i$ the probability that PU $\mbox{pu}_i$ will make a transition from state $j$ to state $k$.
Because the frame length is bounded, $P_{i}(0|l_i)=1$, $\forall i\in\{0,1,...,K_p-1\}$. In addition, $P_{i}(k|j) = 0$ $\forall j,k$ such that $0\leq j<l_i$, $0<k\leq l_i$ and $k\neq j+1$.
In the simulations we set $l_0=3, l_1 = 4, l_2 = 4$ , and $l_3 = 5$. A diagram which illustrates the state transition probabilities for $\mbox{pu}_0$ is depicted in Fig. \ref{fig:markov_chain}.
The transition probabilities for all PUs are summarized in Table \RNum{2}.
\begin{table}[htbp]
\caption{PUs state transition probabilities}
\begin{center}
\begin{tabular}{c||c c c c c c} 
  & $P_i(0|0)$ & $P_i(0|1)$ & $P_i(0|2)$ & $P_i(0|3)$ & $P_i(0|4)$ & $P_i(0|5)$ \\ [0.5ex] 
 \hline\hline
 $i=0$ & 0.1 & 0.1 & 0.15 & 1 & - & - \\ 
 \hline
 $i=1$ & 0.2 & 0.2 & 0.1 & 0.2 & 1 & - \\
 \hline
 $i=2$ & 0.15 & 0.18 & 0.3 & 0.1 & 1 & - \\
 \hline
 $i=3$ & 0.28 & 0.2 & 0.02 & 0.15 & 0.01 & 1 \\
 \hline

\end{tabular}
\end{center}
\end{table}
In the following simulations, we consider three scenarios:
\begin{itemize}
    \item In the first scenario, referred to as Scenario 1, we set the access policy of PUs such that each PU can access a single, fixed, pre-determined channel at each time step, i.e., $\mbox{pu}_i$ can access only $\mbox{ch}_i$, whenever it needs to transmit.
    \item In the second scenario, referred to as Scenario 2, we set the PUs access policy as follows:
    \begin{itemize}
        \item Once a PU begins transmitting at a given channel, it will transmit the entire frame over that channel, e.g., if channel $\mbox{ch}_j$ was allocated to $\mbox{pu}_i$ when $m_i=1$, then channel $\mbox{ch}_j$ will be allocated to $\mbox{pu}_i$ until $m_i=0$, at which this channel allocation is determined.
        \item If a new PU, e.g., $\mbox{pu}_i$ begins to transmit at a given time step ($m_i=1$), it will use channel $\mbox{ch}_k$ for transmission where $\mbox{ch}_k$ is the available channel with the minimal index $k$. This can be justified by an ordering of channels according to some measure of quality, e.g., SNR, where a channel with a larger noise power is assigned a smaller index.
        \item If two PUs or more begin to transmit at a given time step, the PU with the minimal index will use the available channel with the minimal index for transmission. For example, if at any time step, both $\mbox{pu}_i$, and $\mbox{pu}_j$ ($i<j$) begin to transmit, and $ch_k$, $ch_l$ ($k<l$) are available, then $\mbox{pu}_i$ will transmit on $ch_k$, and $\mbox{pu}_j$ will transmit on $ch_l$. This represents a preference assignment where the preferred user has a larger index.
    \end{itemize}

\begin{figure}
  \centering
  \includegraphics[width=0.9\textwidth]{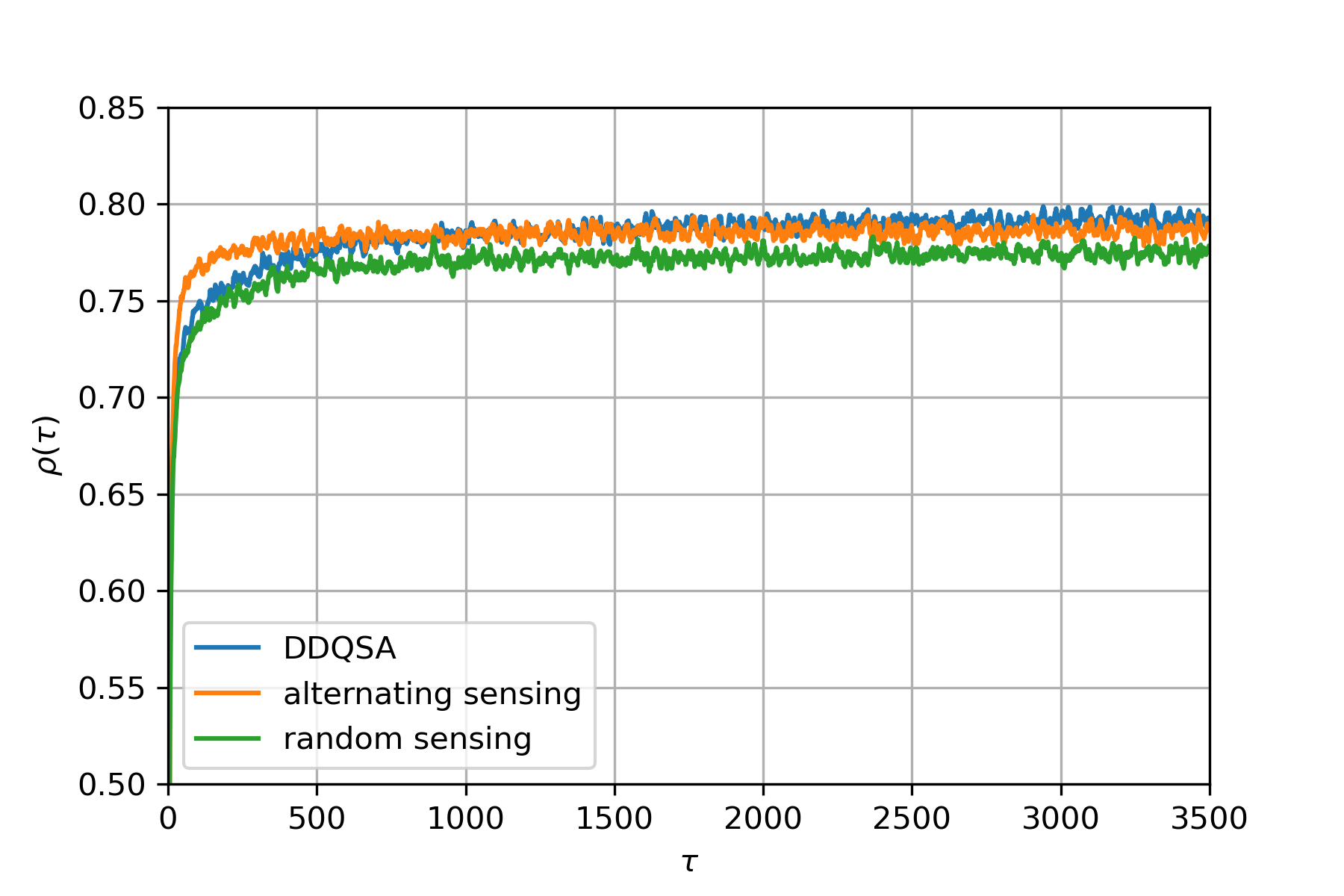}
  \caption{Relative throughputs of the different algorithms for Scenario 1}
  \label{fig:scenario1}
\end{figure}

\begin{figure}
  \centering
  \includegraphics[width=0.9\textwidth]{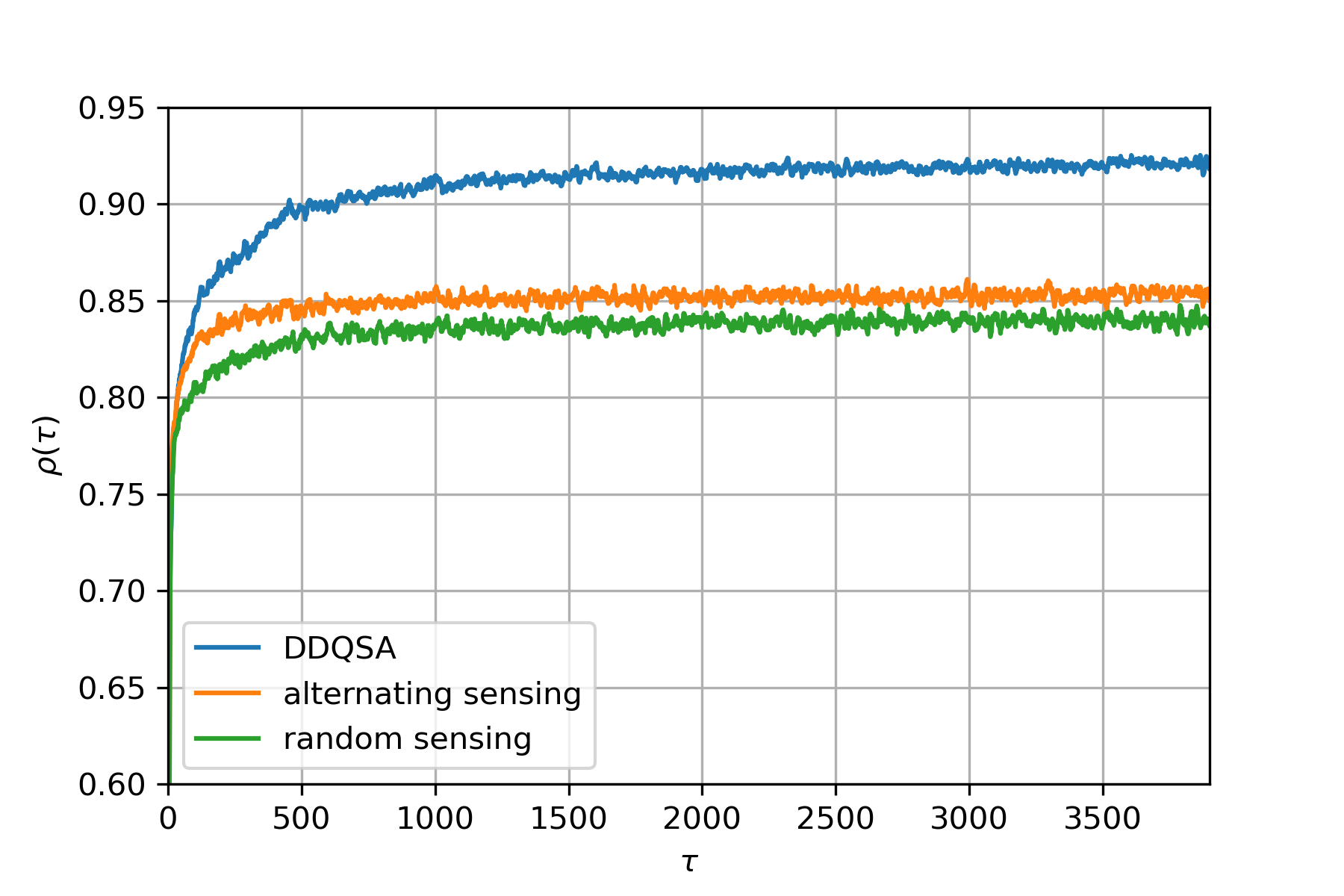}
  \caption{Relative throughputs of the different algorithms for Scenario 2}
  \label{fig:scenario2}
\end{figure}

\begin{figure}
  \centering
  \includegraphics[width=0.9\textwidth]{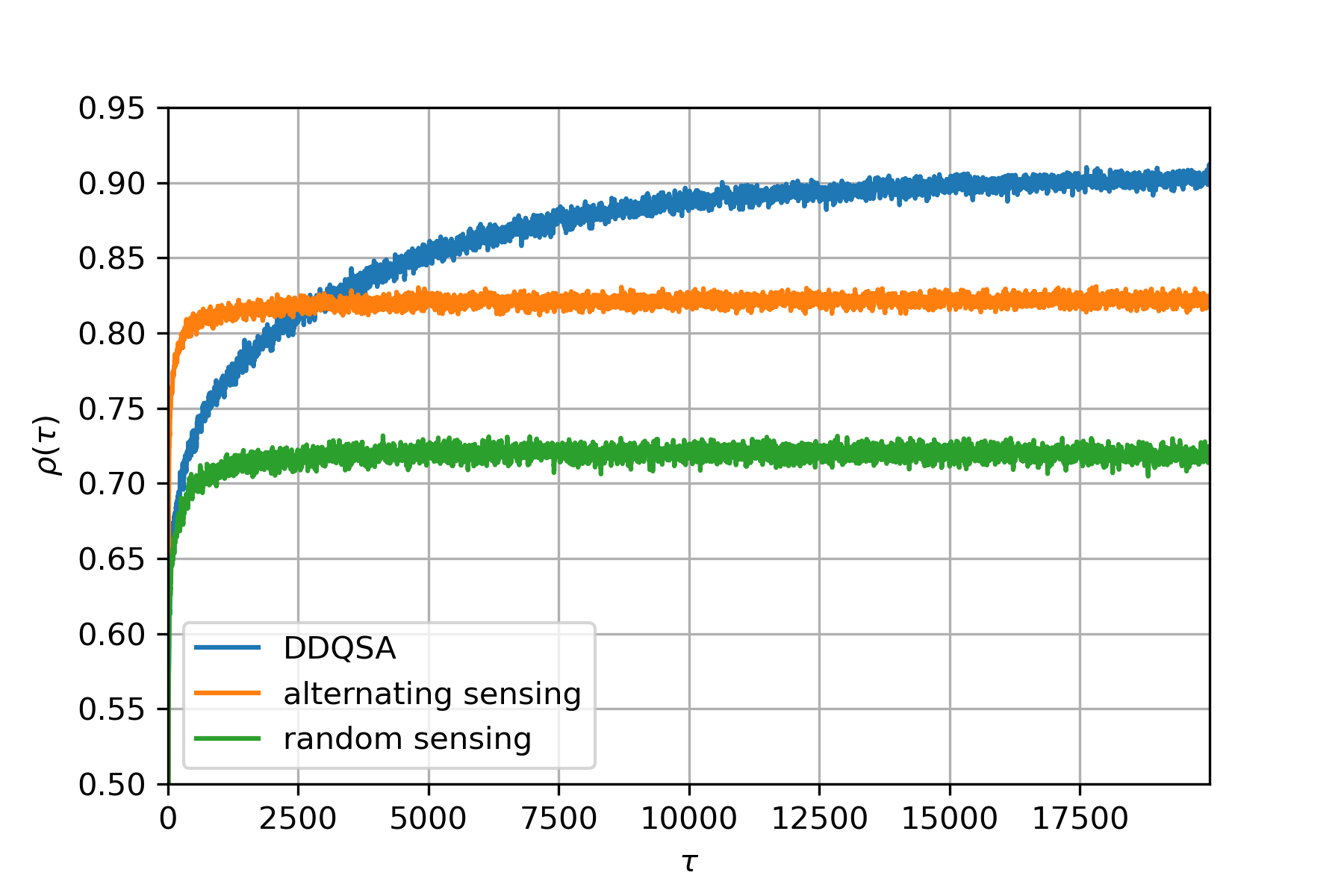}
  \caption{Relative throughputs of the different algorithms for Scenario 3}
  \label{fig:scenario3}
\end{figure}

    \item In the third scenario, referred to as Scenario 3, the PUs follow the same policy as described for Scenario 2, but at every even time step, the channels are flipped, i.e., every 2 time steps, $\mbox{ch}_0$ will switch with $\mbox{ch}_3$, and $\mbox{ch}_1$ will switch with $\mbox{ch}_2$, which corresponds to a frequency hopping network.
\end{itemize}\par
For evaluating the throughput of the random access algorithm, recall that it does not apply any learning process, hence its throughput is evaluated by simply making random access decisions for $1.5\cdot 10^6$ time steps and then averaging the resulting throughput. It follows that the throughput of the random access algorithm is the same for all 3 scenarios described in this subsection, and is approximately $\rho_{RA}(\tau) \approx 0.415$, $\forall\tau\in\mathds{N}$.\par
Fig. \ref{fig:scenario1} depicts the relative throughputs of the different algorithms for Scenario 1. It can be observed from the figure that the DDQSA algorithm achieves the best asymptotic performance, and that the asymptotic throughput achieved by applying alternating sensing is slightly lower than the throughput of the DDQSA algorithm in this case. This result suggests that when each PU accesses a single fixed channel, a near-optimal sensing policy is to sense the subsets of channels alternatingly.
Fig. \ref{fig:scenario2}, depicts the relative throughputs of the different algorithms for Scenario 2. Observe that for this scenario, the throughput achieved by the DDQSA algorithm is significantly superior to that achieved by the other three algorithms. Among the other three referenced algorithms, the best asymptotic relative throughput is achieved by the alternating sensing algorithm, which is about $0.85$, whereas the throughput achieved by the DDQSA algorithm is approximately $0.92$. This clearly demonstrates that the DDQSA is able to learn a sensing policy and correspondence access policy which improves the throughput.
Lastly, Fig. \ref{fig:scenario3}, depicts the relative throughputs for Scenario 3. It is observed again that DDQSA achieves the best throughput asymptotically. Note that in this case the convergence rate is slower, since the PUs have a complex access policy as the states of the network correspond to the state of Scenario 2 for two consecutive time steps and then, for the next two time steps the states are a mirror image of the states of the network in Scenario 2, and so on. This behavior requires from the agent more interactions with the environment in order to learn (near-)optimal policies.

\subsection{Implementations Aspects}

In the above simulations and in the design of the DDQSA algorithm, it was assumed that the SU accesses channel at each time step and thus receives an ACK/NACK signal at each time step. As in practical scenarios a node may also have idle times, we note that this assumption can be easily relaxed, requiring only some minor changes in Algorithm~\ref{alg:my_alg}. To accommodate the fact that the SU may have idle times, we assume that the SU accesses the channel with probability $P_{ac}<1$, and does not access any channel with probability $P_{Nac}=1-P_{ac}>0$. 
The changes in Algorithm~\ref{alg:my_alg} required to deal with the case when $P_{ac}<1$ are as follows:
\begin{itemize}
    \item In line 6 of Algorithm~\ref{alg:my_alg}, the $\epsilon$ should decay as $\epsilon = \frac{1}{1+0.01\cdot \tilde{t}}$, where $\tilde{t}\in\mathds{N}$ is a counter of the time steps in which the SU transmits.
    \item In line 8, the agent executes both actions $a_s(t)$ and $a_{ac}(t)$ when the SU needs to transmit at time step $t+1$, and executes only action $a_s(t)$ when the SU has nothing to transmit at time step $t+1$.
    \item When the SU has nothing to transmit at time step $t+1$, then lines 9 and 10 in Algorithm~\ref{alg:my_alg} should be skipped since the agent will not receive a reward at time step $t+1$.
\end{itemize}\par
\begin{figure}
  \centering
  \includegraphics[width=0.9\textwidth]{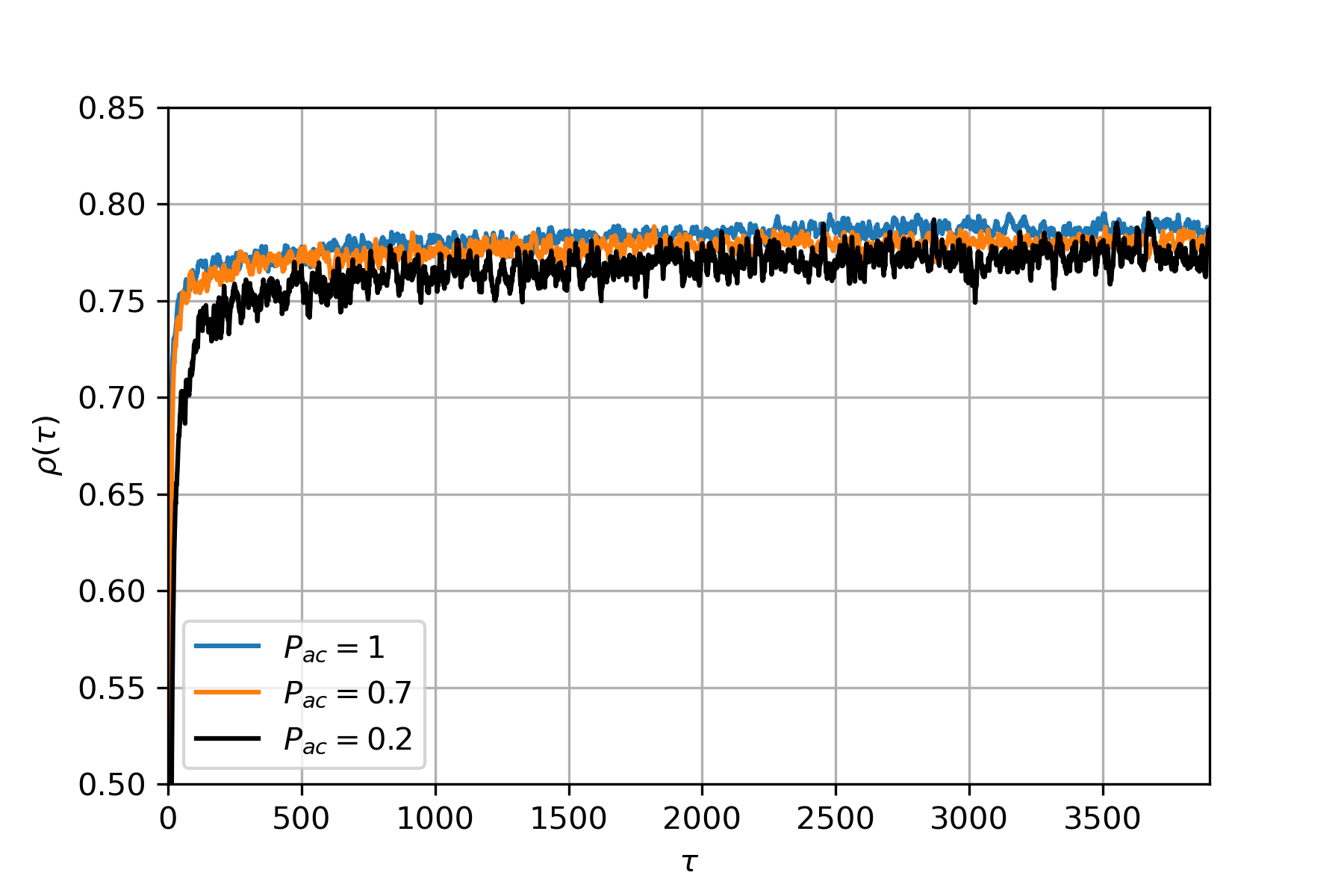}
  \caption{Relative throughputs for the cyclic PU network for different access probabilities}
  \label{fig:su_not_always}
\end{figure}

In Fig. \ref{fig:su_not_always}, we compared the relative throughput of the DDQSA algorithm for the cyclic PU network for the situations in which the SU transmits at every time step with the relative throughput of the modified version of the DDQSA algorithm described above when the SU transmits with probability $P_{ac}=0.7$, and with probability $P_{ac}=0.2$.
From Fig. \ref{fig:su_not_always}, it is observed that the relative throughput when $P_{ac}=0.7$ and $P_{ac}=0.2$ are very close to the relative throughput when the SU transmits at every time step. When the SU transmits with a small probability, e.g., $P_{ac}=0.2$, $\rho(\tau)$ is noisier because $\rho(\tau)$ involves an averaging operation over 100 time-steps. In addition, it is observed that as $P_{ac}$ decreases, the convergence rate becomes slower due to the fact that the replay buffer is loaded less frequently when the SU transmits infrequently.

\section{Conclusion}

We considered the DSA problem, where multiple PUs access a network according to a predetermined scheme and a cognitive SU, which has no prior knowledge about the PUs dynamics and the access policy they use, and attempts to access the channel. In order to successfully transmit, the SU estimates the indices of free channels. To that aim, the SU is capable of sensing a subset of the available channels at each time step, due to sensing bandwidth limitations. To identify the SU policy which maximizes its throughput, we developed a novel DDQSA algorithm, which aims to determine the best sensing strategy and the corresponding best access strategy, based on past observations collected by the SU via online learning. We compared the throughput of the proposed DDQSA algorithm with that of three other algorithms which use pre-determined sensing or access policies for four different scenarios. The results showed that DDQSA outperforms the baseline algorithms in all cases. 
Moreover, for Scenario 1, in which the PUs use a cyclic access policy, we analytically derived the optimal sensing and access policies and the corresponding maximal throughput. In this scenario, the throughput of DDQSA is very close to the optimal throughput and is significantly higher than the throughputs achieved by the other three algorithms. Finally, we demonstrated that a modified version of the suggested DDQSA algorithm can be applied to more practical scenarios in which the SU does not transmit at every time step.

These results clearly demonstrate the ability of DDQSA to learn near-optimal policies and the overall superiority of the proposed approach over existing methods.

\bibliographystyle{IEEEtran}
\bibliography{YBpaper}

 \end{document}